\documentstyle[twoside,fleqn,espcrc2,psfig]{article}
\newcommand{\beq}{\begin{equation}}
\newcommand{\eeq}{\end{equation}}
\newcommand{\beqa}{\begin{eqnarray}}
\newcommand{\eeqa}{\end{eqnarray}}
\newcommand{\beqan}{\begin{eqnarray*}}
\newcommand{\eeqan}{\end{eqnarray*}}

\newcommand{\AmS}{{\protect\the\textfont2
  A\kern-.1667em\lower.5ex\hbox{M}\kern-.125emS}}

\title{Fermion Decoupling and the Axial Anomaly on the Lattice}

\author{H. Banerjee\address{S.N.Bose National Centre for Basic Sciences,
JD Block, Salt Lake, Calcutta 700 091, India}
and
Asit K. De\address{Saha Institute of Nuclear Physics, 1/AF
Salt Lake, Calcutta 700064, India}\thanks{Presenter at Lattice'99.}}

\begin{document}

\begin{abstract}
In the axial Ward identity of lattice QED we show that in the limit of
infinite fermion mass $m$ the pseudoscalar density term exactly cancels the
Adler-Bell-Jackiw (ABJ) anomaly. Using this result we calculate the U(1) axial
anomaly in a non-abelian gauge theory.
\end{abstract}
\maketitle

\section{Introduction.}
Wilson fermions break chiral symmetry explicitly. Explicit
breaking of chiral symmetry is necessary to generate,
from the lattice regulated model, the U(1) axial anomaly in
the continuum limit for a vectorlike gauge theory \cite{KaSm81}.

To examine the role of the underlying lattice fermion model in generating
the ABJ anomaly a convenient and transparent starting point is the
condition, in this context,
for decoupling of the fermion in the large mass limit from the background
gauge field \cite{Apple},
\begin{eqnarray}
\langle \Delta_\mu J_{\mu 5}(x)\rangle_{a=0}
&=&
2im \;\; \langle \overline{\psi}_x\gamma_5\psi_x\rangle_{a=0} \nonumber \\
&-&\!\!\!\lim_{m\rightarrow
\infty}\left[2im\langle\overline{\psi}_x\gamma_5\psi_x
\rangle_{a=0}\right] \label{decoupling} \end{eqnarray}
where $a$ is the lattice constant. One recognises Eq.(\ref{decoupling}) as
the Adler condition \cite{Adler} which states that the triangle graph
amplitude should vanish in the limit as the mass of the loop fermion becomes
infinite. To establish that the decoupling condition is indeed equivalent to
the axial Ward identity one needs the supplementary relation
\begin{equation}
\lim_{m\rightarrow \infty}\!\!\left[2im\!\langle\overline{\psi}_x
\gamma_5\psi_x\rangle_{a=0}\right]\! = \!\frac{ig^2}{16\pi^2}
\epsilon_{\mu\nu\lambda\rho}{\rm tr}F_{\mu\nu}F_{\lambda\rho} \label{anomaly1}
\end{equation}
where $F_{\mu\nu}$ is the gauge field tensor.

Our derivation of Eqs.(\ref{decoupling}) and (\ref{anomaly1})
in lattice QED demonstrates that as long as the underlying lattice fermion
model removes doubling completely and is gauge-invariant and local, the ABJ
anomaly is generated without reference to the specific form of the irrelevant
term. In non-abelian gauge theories on lattice Eq.(\ref{decoupling}) provides,
as we shall see, a simple recipe for deriving the U(1) axial
anomaly.

\section{Decoupling in QED.}
The key to our analysis is the Rosenberg \cite{Rosen} tensor decomposition
of the amplitude of the triangle diagrams (i) and (ii) in continuum QED for
axial current $j_{\lambda 5}(x)$ to emit two photons with momenta and
polarisation ($p,\mu$) and ($k,\nu$):
\begin{eqnarray}
&T^{(i+ii)}_{\lambda\mu\nu}&\!\!\! = \;\;
\epsilon_{\lambda\mu\nu\alpha}  k_\alpha A(p,k,m) \nonumber \\
&+& \!\!\!\!\!\!\!\!\!\!
\epsilon_{\lambda\nu\alpha\beta} p_\alpha k_\beta [p_\mu B(p,k,m)
+ k_\mu C(p,k,m)] \nonumber \\
&+& \!\!\!\!\!\!\!\!\!\![(k,\nu) \leftrightarrow (p,\mu)].
\label{rosen}
\end{eqnarray}

Gauge invariance relates the Rosenberg form factors $B$ and $C$ to $A$.
\begin{equation}
A(p,k,m)=p^2B(p,k,m)+p.k\; C(p,k,m).\label{gi}
\end{equation}

The form factors $B$ and $C$ are of mass
dimension -2, and, therefore, must vanish as $m^{-2}$ for large fermion
mass. Gauge invariance then guarantees that

\beqa
& &\lim_{m\rightarrow\infty} (p+k)_\lambda
T^{(i+j)}_{\lambda\mu\nu} = \nonumber \\
\!\!\!\!\!\!\!\!\!\!\!&-&\!\!\!\!\! \epsilon_{\mu\nu\alpha\beta}
p_\alpha k_\beta \lim_{m\rightarrow\infty}[A(p,k,m)+A(k,p,m)]  \label{anomlim}
\\ &=& 0 ,\label{proof}
\eeqa
which is the basis of Eq.(1). In the above, (\ref{proof}) follows from
(\ref{anomlim}) because of (\ref{gi}) and the asymptotic behavior of $B$ and
$C$.

On lattice, the decoupling condition (\ref{proof}) should be realised in the
continuum limit irrespective of the underlying model for fermion as long as
it is free from doublers and local. The form factors $B$ and $C$ which are
highly convergent amplitudes must coincide with their respective expressions
in the continuum in all {\em legitimate} lattice models. Residual model
dependence, if any, can appear only in the form factor $A$ because of
potential logarithmic divergence. This, however, is ruled out by the gauge
invariance constraint (\ref{gi}).

\begin{figure}[t]
\begin{center}
\vspace{-1.5cm}
\hspace*{-1.0cm}\psfig{figure=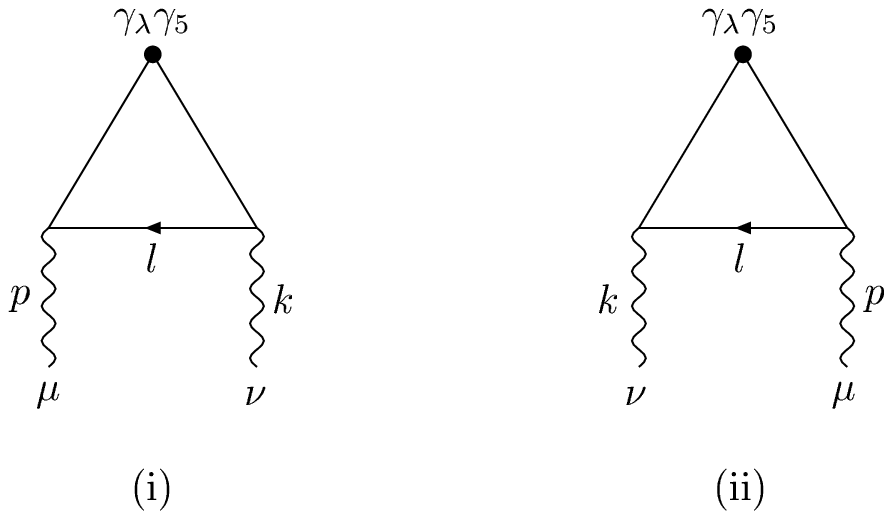,width=13.0cm,height=16.0cm}
\vspace{-1.0cm}
\end{center}
\vspace{-11.0cm}
\end{figure}

In lattice QED with Wilson fermions, 
the Feynman amplitudes corresponding to the two diagrams (i) and (ii) are :
\beqa
[T^{(i+ii)}_{\lambda\mu\nu}]_a \!\!\!\!\!&=&\!\!\!\!\! -g^2 \!\!\!
                     \int_{-\frac{\pi}{a}}^{\frac{\pi}{a}}
\!\!\frac{d^4l}{(2\pi)^4}       {\rm Tr}\Bigg[\gamma_\lambda\gamma_5 \cos
a(l\!+\!\!\frac{k-p}{2})_\lambda  \nonumber \\
& & S(l-p) V_\mu(l-p,l) S(l)  \nonumber \\
& & V_\nu(l,l+k) S(l+k) \nonumber \\
& &
+ (k,\nu \leftrightarrow p,\mu)\Bigg],    
\label{ampli}
\eeqa
with the fermion
propagator $S(l)$ and the one-photon vertex $V_\mu(p,q)$ given by
\beqa
S(l)\!\!\! &=& \!\!\!\!\!\!\left[ \sum_\mu
\gamma_\mu \frac{\sin al_\mu}{a}\! + \!\frac{r}{a} \sum_\mu (1-\cos
al_\mu)\!+\!m \right]^{-1}\nonumber \\
V_\mu \!\!\!\!\!\!&=& \!\!\!\!\!\gamma_\mu \cos
\frac{a}{2}(p+q)_\mu + r \sin \frac{a}{2}(p+q)_\mu. \nonumber
\eeqa
where $r$ is the Wilson parameter. {\em We have the same $\gamma$-matrix
convention as in} \cite{KaSm81}.
On lattice there are four additional diagrams with {\em irrelevant} vertices.
As will be evident from the following, they do not contribute in the continuum
limit.

The lattice amplitude (\ref{ampli}) is superficially linearly divergent.
However, the leading term, obtained by setting the
external momenta $p, \;k=0$ is odd in the loop momentum $l$ and
vanishes due to symmetric integration. The amplitude, therfore, vanishes at 
least linearly in external momenta as indeed the Rosenberg decomposition 
suggests and, furthermore, the effective divergence is at most logarithmic.

Our strategy is to consider the derivative of (\ref{ampli}) with respect to the
fermion mass $m$ rather than the external momenta $p,\;k$ as is common practice
\beq
[R^{(i)}_{\lambda\mu\nu}]_a\equiv
\frac{d}{dm}[T^{(i)}_{\lambda\mu\nu}]_a . \label{R}
\eeq  
Lattice power counting gives a negative integer for the effective degree of
divergence of $[R^{(i)}_{\lambda\mu\nu}]_a$. 
One can, therefore, take, thanks to the Reisz theorem \cite{Reisz},
the continuuum limit of the integrands and evaluate the loop integrals in
the entire phase space $-\infty \le l_\mu \le \infty$ as in the continuum.
In the continuum limit, amplitudes of only two
diagrams (i) and (ii) survive and amplitudes with {\em irrelevant} vertices 
vanish. The amplitudes
$[R^{(i)}_{\lambda\mu\nu}]_{a=0}$ and $[R^{(ii)}_{\lambda\mu\nu}]_{a=0}$ 
are individually Bose-symmetric and hence gauge-invariant

The Rosenberg tensor decomposition is

$[R^{(i+ii)}_{\lambda\mu\nu}]_{a=0}=$
\beqa
& &4g^2 m \int^\infty_{-\infty} \frac{d^4l}{(2\pi)^4}
\Bigg[{\rm Tr}(\gamma_5 \gamma_\lambda \gamma_\mu \gamma_\nu p\!\!/)\nonumber
\\
& &\left(\frac{1}{D}(1+\frac{k^2}{d_3})-\frac{1}{d_1d^2_3}\right) \nonumber \\
&+&\!\!\!\!\!{\rm Tr}(\gamma_5\gamma_\lambda\gamma_\nu p\!\!/ k\!\!\!/)
\frac{2(l_\mu-p_\mu)}{Dd_1} + (p,\mu \leftrightarrow k,\nu)\Bigg]\label{Rosen}
\eeqa
where
$D=d_1d_2d_3 \;\;\; {\rm and}$\\
$d_1\equiv (l-p)^2+m^2,\; d_2\equiv l^2+m^2,\; d_3\equiv (l+k)^2+m^2$.

The four-divergence of the amplitude for the axial vector current is to be
obtained from
\beq
[(p+k)_\lambda
R^{(i+ii)}_{\lambda\mu\nu}]_{a=0} =\frac{d}{dm}[(p+k)_\lambda
T^{(i+ii)}_{\lambda\mu\nu}]_{a=0} \nonumber
\eeq
\beq
= -\frac{1}{\pi^2}\epsilon_{\mu\nu\alpha\beta}
p_\alpha
k_\beta \frac{d}{dm}\int_{0\le s+t\le 1}\frac{m^2}{c^2+m^2} ds\;dt,
\eeq
\beq
\mbox{with}\;\;
c^2\equiv s(1-s)p^2 + t(1-t)k^2 +2st\;p.k.
\eeq

The Adler condition (1) determines the {\em constant of integration}\\

\noindent $[(p+k)_\lambda T^{(i+ii)}_{\lambda\mu\nu}]_{a=0}
= $
\beq
-\frac{1}{\pi^2}\epsilon_{\mu\nu\alpha\beta}p_\alpha k_\beta
\left[\int \frac{m^2}{c^2+m^2}ds\;dt -\frac{1}{2}\right] \label{div}.
\eeq 

The ABJ anomaly is identified as the $m=0$ limit of the right hand side of 
(\ref{div}):
\beq
{\rm ABJ\;\; anomaly} = \frac{1}{2\pi^2}\epsilon_{\mu\nu\alpha\beta}p_\alpha 
k_\beta \label{anomaly}
\eeq

\noindent
\section{U(1) axial anomaly in non-abelian gauge theories.}
The representation motivated by the decoupling condition (1):
\beqa
\noindent
\!\!\!\!\!\!&&\lim_{m\rightarrow
\infty}\left[2im\langle\overline\psi_x\gamma_5\psi_x\rangle_{a=0}\right]
= \nonumber \\
\!\!\!\!\!\!&&\lim_{m\rightarrow \infty}\left[2im\langle x|{\rm
Tr}\gamma_5(D\!\!\!\!/ +W+m)^{-1)}|x\rangle_{a=0}\right] \label{PV}
\eeqa
constitutes the starting point of our calculation of the axial anomaly in
non-Abelian theories, {\em e.g.}, lattice QCD. The Dirac operator 
$D\!\!\!\!/$ and the Wilson term $W$ are given by
\beqa
D_\lambda & \equiv & \frac{1}{2ia}\left(e^{ip_\lambda a} U_\lambda
                    -U_\lambda^\dagger e^{-ip_\lambda a}\right) \\ \nonumber
W &\equiv & \frac{r}{2a} \sum_\lambda \left(2-e^{ip\lambda a} U_\lambda
                     -U^\dagger_\lambda e^{-ip_\lambda a}\right)
\eeqa
where $U_\lambda\equiv exp(iagA_\lambda)$ is the link variable with
$A_\lambda \equiv t^a A^a_\lambda$ the gauge potential and $t^a$ the
generators of $SU(N)$.

Our strategy is to develop the Green
function for lattice fermion in a perturbative series:\\

\noindent $
(D\!\!\!\!/+W+m)^{-1} = (-D\!\!\!\!/+W+m)G, \;\;\mbox{with} $
\vspace{-0.2cm}
\beqa
G &=&\left(-D\!\!\!\!/^2+(W+m)^2+[D\!\!\!\!/,W]\right)^{-1}
\nonumber \\
&=& G_0 -gG_0VG_0 +g^2 G_0VG_0VG_0+ ... \label{perturb}
\eeqa
where the {\em free} part $ G_0=$\\
$\left[\sum\frac{\sin^2ap_\mu}{a^2}+ \left(\frac{r}{a}\sum_\mu(1-\cos
ap_\mu)+m\right)^2\right]^{-1}$
is of Reisz degree $-2$ and has the expected continuum limit
$(p^2+m^2)^{-1}$.

The potential $gV$ has three pieces
\beq
gV = gV_0 + gV_1 + gV_2
\eeq
of which the first piece $gV_0$ is independent of $\gamma$-matrices, has
Reisz degree $+1$ and non-vanishing continuum limit. The pieces $gV_1$ and
$gV_2$ contain $\gamma$-matrices and each has Reisz degree zero. The
continuum limit of $gV_1$ vanishes
\beq
(gV_1)_{a=0} = [D\!\!\!\!/,W]_{a=0} =0,
\eeq
whereas, 
\beq
(gV_2)_{a=0}= \frac{i}{2} \sigma_{\mu\nu} \left[D_\mu,D_\nu\right]_{a=0}
           = -\frac{i}{2} \sigma_{\mu\nu} F_{\mu\nu}
\eeq
where $F_{\mu\nu}$ is the field tensor in the continuum.

The first two terms of the perturbative series
(\ref{perturb}) do not contribute simply
because they do not have enough $\gamma$-matrices to give non-vanishing
Dirac trace. Reisz power counting for the second and higher order terms
in (\ref{perturb}) all
give negative integers. One can now use the Reisz theorem and take the
continuum limit of the integrands in all these terms. Anomaly is thus given
by the term which survive in the large mass limit in the continuum
\beqa
\!\!\!\!\!\!\!\!\!&&
-\lim_{m\rightarrow \infty} \left[ 2img^2\langle x|{\rm Tr} \gamma_5 G_0
V_2 G_0 V_2 G_0|x\rangle_{a=0}\right] \nonumber \\
&=&
-\;\frac{ig^2}{16\pi^2}\epsilon_{\lambda\rho\mu\nu} {\rm tr}
F_{\lambda\rho}(x) F_{\mu\nu}(x) \label{final}
\eeqa
where `tr' now denotes trace over internal symmetry indices. Note that the
final result (\ref{final}) is local, all nonlocalities disappearing in the
large $m$ limit, as do all higher order terms in the perturbative series
(\ref{perturb}).


\begin{thebibliography}{9}

\bibitem{KaSm81} L.H. Karsten and J. Smit, Nucl. Phys. B127 (1981), 103.

\bibitem{Apple} T. Appelquist and J. Carazzone, Phys. Rev. D 11 (1975) 2856.

\bibitem{Adler} Appendix A of S.L. Adler, Phys. Rev. 177 (1969) 2426;
		K. Fujikawa, Z. Phys. C 25 (1984) 179.

\bibitem{Rosen} L. Rosenberg, Phys. Rev. 129 (1963) 2786.

\bibitem{Reisz} T. Reisz, Comm. Math. Phys. 116(1988)81;
\end{thebibliography}
\end{document}